\documentclass[pra,twocolumn,,superscriptaddress,showpacs]{revtex4}

\usepackage{amssymb,amsmath,epsfig}

\def\eq#1{Eq.~(\ref{#1})}
\def\fig#1{Fig.~\ref{#1}}

\begin{document}

\title{Band Structure, Phase transitions and Semiconductor Analogs in One-Dimensional Solid Light Systems}

\author{James Quach} \email{jamesq@unimelb.edu.au}
\affiliation{School of Physics, The University of Melbourne, Victoria 3010, Australia}
\affiliation{Centre for Quantum Computer Technology, School of Physics, The University of Melbourne, Victoria 3010, Australia}

\author{Melissa I. Makin}
\affiliation{School of Physics, The University of Melbourne, Victoria 3010, Australia}
\affiliation{Centre for Quantum Computer Technology, School of Physics, The University of Melbourne, Victoria 3010, Australia}

\author{Chun-Hsu Su}
\affiliation{School of Physics, The University of Melbourne, Victoria 3010, Australia}
\affiliation{Centre for Quantum Computer Technology, School of Physics, The University of Melbourne, Victoria 3010, Australia}

\author{Andrew D. Greentree}
\affiliation{School of Physics, The University of Melbourne, Victoria 3010, Australia}

\author{Lloyd C.L. Hollenberg}
\affiliation{School of Physics, The University of Melbourne, Victoria 3010, Australia}
\affiliation{Centre for Quantum Computer Technology, School of Physics, The University of Melbourne, Victoria 3010, Australia}

\begin{abstract}

The conjunction of atom-cavity physics and photonic structures (``solid light'' systems) offers new opportunities in terms of more device functionality and the probing of designed emulators of condensed matter systems. By analogy to the canonical one-electron approximation of solid state physics, we propose a one-polariton approximation to study these systems. Using this approximation we apply Bloch states to the uniformly tuned Jaynes-Cummings-Hubbard model to analytically determine the energy band structure. By analyzing the response of the band structure to local atom-cavity control we explore its application as a quantum simulator and show phase transition features absent in mean field theory. Using this novel approach for solid light systems we extend the analysis to include detuning impurities to show the solid light analogy of the semiconductor. This investigation also shows new features with no semiconductor analog.

\end{abstract}


\maketitle

\section{Introduction}

Quantum properties of strongly correlated many-particle systems are intensive to compute and experimentally hard to access in conventional solid state materials. Because of this, artificial tunable systems to simulate highly correlated many-body dynamics have been proposed. Early attempts at these so called \emph{quantum simulators} used arrays of Josephson junctions \cite{vanderzant92,vanderzant96} to reproduce strong many-body bosonic interactions. Cold atoms trapped in optical lattices \cite{Jaksch97} now constitute an important experimental platform for quantum simulation. Recently, arrays of coupled atom-photon cavities have been proposed \cite{hartmann06,greentree06,angelakis07} as another highly flexible platform for quantum simulation. This has opened up exciting new possibilities for combining atom-cavity and photonic structures. A two-level atom in an optical cavity is described by the well known Jaynes-Cummings (JC) \cite{jaynes63} model that gives rise to nonlinear photon-photon interactions including photon blockade \cite{Tian92,Imamoglu97,rebic02,birnbaum05}. By considering a network of evanescently-coupled JC systems, a tight binding model can be developed, which is the Jaynes-Cummings-Hubbard (JCH) model \cite{greentree06,angelakis07,makin07,Schmidt09}. In this paper, we consider this particular model and by determination of its band structure in response to local atom-cavity control we explore the application of the system as a quantum simulator.

\begin{figure}
	\includegraphics[width=80mm]{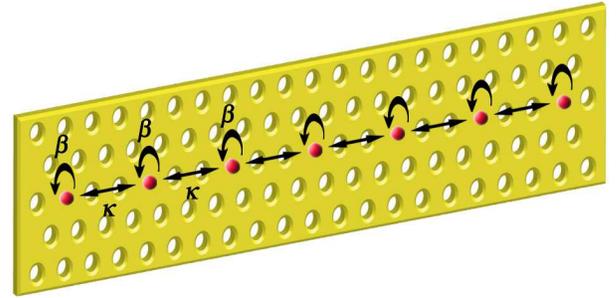}
	\caption{(Color online)  A possible realisation of a one dimensional Jaynes-Cummings-Hubbard model. Here holes are drilled into a thin membrane and lattice defects serve as the optical cavities housing two-level atoms. The cavities are coupled evanescently with single-photon hopping rate $\kappa$ and the atom is coupled to the local cavity mode with coupling strength $\beta$.}
	\label{fig:JCH_realisation}
\end{figure}

A number of realizations of the JCH model are possible including photonic bandgap (PBG) structures with single atoms \cite{greentree06} or small ensembles \cite{Na08}, arrays of superconducting strip-line cavities \cite{wallraff04}, cold atoms connected via optical fibre interconnects \cite{Trupke05}, plasmonics \cite{Chang06}, or linear ion traps \cite{Ivanov09,Mering09}. For clarity we focus on the conceptually simple PBG implementation. A photonic crystal is realized by a periodic modulation of refractive index that is arranged so as to have an optical bandgap i.e. a range of photon energies that can not propagate through the medium \cite{Joannopoulos95}. Typical implementations are arrays of holes drilled in a high refractive index membrane \cite{Yamaguchi09}. Lattice defects, i.e. where holes are not drilled, act as optical cavities into which a two-level atom can be placed as illustrated in \fig{fig:JCH_realisation}. If the membrane was fabricated in diamond, the two-state systems could be ion-implanted NV centers, and there exists suitable designs for such cavities \cite{Bayn08,Tomljenovic-Hanic08}. The photon-atom coupling provides the mechanism by which the polaritons strongly interact.

An infinite JCH lattice exhibits a quantum phase transition analogous to the insulator-superfluid transition of the Bose-Hubbard model. This has been confirmed by several methods including mean-field (MF) \cite{greentree06,Koch09}, density matrix renormalization group \cite{Rossini07}, variational cluster \cite{aichhorn07}, and linked-cluster expansion \cite{Schmidt09}, and quantum Monte Carlo \cite{Pippan09}. Zhao et. al \cite{Zhao08} have also argued that the JCH phase transition belongs to a different universality class from the Bose-Hubbard model. The signature of transition has also been identified in finite cavity simulations \cite{makin07}. The term ``solid light'' was so coined because of the presence of the Mott insulation (MI) phase in the systems. 

The macroscopic to mesoscopic size of conventional solid state systems, has meant that it has been natural in solid state physics to assume infinite lattices. Traditional cavity QED systems have been restricted to few (usually one) cavities, however this is changing due to advances in fabrication \cite{Notomi08}. These new opportunites reveal a parameter space where condensed-matter concepts can be usefully applied to quantum optical systems, which in return act as quantum simulators to study non-trivial condensed matter systems.

Within this context, we consider the case of a one-dimensional (1D) infinite chain of cavities. By analogy with the canonical one-electron approximation in solid state physics, we discuss the \textit{one-polariton approximation} in a solid light context. This work informs and extends the literature on 1D coupled-cavity waveguide photon propagation \cite{Paternostro09,Lu09,Makin09}. In particular, following a review of the JCH model in Sec.~\ref{sec:The Jaynes-Cummings-Hubbard Model}, we apply Bloch states to analytically develop the band structure of solid light, and we ascertain the phase transition diagram of Mott lobes and the superfluid (SF) regions in Sec.~\ref{sec:The One Polariton Approximation And Band Structure}. One should note that, while the band structure of PBG systems is well understood, we are exploring the band structure of the defects within the PBG, i.e. bands within the band gap. In Sec.~\ref{sec:Semi-Fluid In One Dimension} we extend the one-polariton analysis to investigate how semiconductor properties can be simulated in the JCH model by embedding detuned atoms in a tuned system, showing a solid light analog of the extrinsic semiconductor.

\section{The Jaynes-Cummings-Hubbard Model}
\label{sec:The Jaynes-Cummings-Hubbard Model}

The Jaynes-Cummings (JC) model \cite{jaynes63} describes a two-level atom interacting with a quantized mode of an optical cavity. It is described by the Hamiltonian ($\hbar=1$),
\begin{equation}
	H^{\rm{JC}} = \epsilon \sigma^+ \sigma^- + \omega a^\dagger a + \beta(\sigma^+ a + \sigma^- a^\dagger),
\label{eq:jaynes_cummings_Hamiltonian_1}
\end{equation}
where $\sigma^+,\sigma^- (a^\dagger,a)$ are the atomic(photonic) raising and lowering operators respectively, $\epsilon$ is the atomic transition energy, $\omega$ is the cavity resonance frequency, and $\beta$ is the atom-photon coupling.

The eigenvectors of the JC Hamiltonian are the dressed states (polaritons),
\begin{equation}
	\vert \pm,n \rangle = \frac{\beta\sqrt{n}\vert g,n \rangle + [-\Delta/2 \pm \chi(n)]\vert e,n-1 \rangle}{\sqrt{2\chi^2(n) \mp \chi(n)\Delta}}~,\forall~n>0,
\label{eq:jc_evectors}
\end{equation}
with corresponding eigenvalues,
\begin{equation}
	 E_{\vert \pm,n \rangle} = n\omega \pm \chi(n) - \Delta/2,
\label{eq:jc_evalues}
\end{equation}
where $\vert g,n \rangle$ and $\vert e,n-1 \rangle$ are the atomic ground and excited states with $n$ excitations, detuning parameter $\Delta \equiv \omega - \epsilon$, and $n$ photon generalized Rabi frequency $\chi(n) = \sqrt{n\beta^2 + \Delta^2/4}$. For $n=0$, the eigenstate is $\vert g,0 \rangle$ with zero eigenvalue.

A feature of the JC model is photonic blockade, whereby absorption of a photon blocks absorption of a subsequent photon because the required transition is detuned from resonance. This effect provides a mechanism by which polaritons can strongly interact. Strong atom-photon coupling is required for photonic-blockade to be observed and ideally very small cavities with very high quality factors, Q, are necessary.

To go beyond one-cavity systems, we consider the Hubbard model \cite{hubbard63} which is one of the simplest of the models describing a system of interacting particles in a lattice. The Hubbard model parameterizes the Hamiltonian by two energy scales, those of inter-site hopping, and on-site repulsion. Quantum phases are seen as the relative strength of these interactions are varied at a defined chemical potential.

Combining the JC and Hubbard models, the JCH model describes an array of Jaynes-Cummings (JC) cavities with Hubbard-like inter-cavity interaction where the JC interactions provides an effective particle-particle interaction. Grouping the JCH Hamiltonian into intra and inter-cavity terms yields,
\begin{equation}
	H = \sum_r{H_r^{\rm{JC}}} - \sum_{\langle r, s \rangle}{\kappa_{rs} a_r^\dagger a_s} - \sum_r{\mu_r N_r},
\label{eq:jaynes_cummings_hubbard_Hamiltonian_1}
\end{equation}
where $r$ and $s$ are cavity site indices, $\mu$ is the chemical potential, and $N$ is the total number of cavity excitations.

The chemical potential energy is the change in mean energy per additional particle, $\mu =\partial E/ \partial N$. While this is typically zero in photonic systems, the nonlinear interactions generated by the JC Hamiltonian makes $\mu$ non-zero \cite{makin07,grochol09}.  One should realize that in our case the chemical potential is not an independent parameter.  In general the cost of adding particles is a function of all of the system parameters, in particular the strength of the atom-photon interactions and the number of excitations. Experimentally, $\mu$ can be determined via spectroscopy analysis of the change in system energy with excitations. However, for easy comparison with the standard phase transition literature, we here treat $\mu$ as an independent parameter.

\section{The One Polariton Approximation And Band Structure}
\label{sec:The One Polariton Approximation And Band Structure}

Because of the strongly interacting nature of the polaritons, the JCH system becomes a complex many-body problem, akin to that of the electronic many-body problem of solid state physics. Complexity quickly increases in difficulty with number of particles making exact solutions difficult to obtain. An approximation which has proven fruitful in describing the electronic properties of solids is the so called one-electron approximation \cite{reitz55}. In this formulation one decouples a single electron from the rest, treating the others as a mean sea of electrons. Thus along with the potential provided by the fixed lattice the single electron also sees a mean potential provided by the sea of charge distribution of the other electrons. Consequently, the total state of the system can be expressed as a linear combination of the states of one electron.

Motivated by the one-electron approximation we investigate the JCH system in a \textit{one-polariton approximation}, describing an extra polariton above a uniformly filled lattice of JC systems. In the photon-blockade regime, we can also consider the concept of a polariton hole, i.e. a cavity with one fewer polariton than in the rest of the lattice. The one-polariton (hole) approximation is most valid in the Mott phase and, as we will show, captures the transition to the superfluid phase, although the approximation will not be valid to describe the SF phase.  The one-polariton (hole) approximation is defined as a reduced basis set,
\begin{equation}
 \{\vert \phi^l_{m\pm1},r \rangle \bigotimes_{s \neq r}{c_l\vert \phi^l_{m},s \rangle}\}~\forall~r,m,
\label{eq:one_photon_basis_1}
\end{equation}
where $r$ and $s$ indicate atomic sites, $\phi^l_m$ represents a site state with $m$ total atomic and photonic excitations, $l$ is indexed over all unique states of $m$ excitations. The repeated index $l$ is summed over all states with $m$ excitations. For example in the bare basis, the set of $l$ would be the ground and excited atomic states. The state co-efficients are such that the normalization condition $\sum_l{c_l^2} = 1$ is satisfied. For the basis set to be most valid in the Mott phase we choose $c_l$ such that in the limit of zero hopping potential (i.e. $\kappa = 0$), each site is a JC eigenstate. In the bare basis $c_l$ are just the co-efficients in the dressed states, i.e.
\begin{align}
	c_g(n) &\equiv \frac{\beta\sqrt{n}}{\sqrt{2\chi^2(n) \mp \chi(n)\Delta}}\\
	c_e(n) &\equiv \frac{-\Delta/2 \pm \chi(n)}{\sqrt{2\chi^2(n) \mp \chi(n)\Delta}}
\end{align}

For notational convenience, in the rest of this paper we will adopt the one-electron convention where the $s$ identical sites of the basis set are implied.

We restrict our attention to the low-temperature limit of a 1D infinite periodic chain of uniform sites, as depicted in \fig{fig:JCH_realisation}. The onsite energy is given by, 
\begin{equation}
	\sum_{s}{H_{rs} \vert \phi,s \rangle = E_r \vert \phi,r \rangle},
\label{eq:onsite_energy_general}
\end{equation}
where  $H_{rs}$ is the Hamiltonian that relates site $r$ to site $s$ and $E_r$ is the onsite energy.

The one-polariton approximation allows us to use the usual ansatz of Bloch's theorem for periodic structures to solve for the onsite eigenenergy. The ansatz is the discretized Bloch state,
\begin{equation}
	\vert \phi,s \rangle = \vert \phi,r \rangle \exp[i \vec{k}\cdot(\vec{d_s}-\vec{d_r})],
\label{eq:discrete_bloch}
\end{equation} 
where $\vec{d_s}$ is the displacement to site $s$, and $\vec{k}$ is the wavevector associated with the crystal momentum. We normalize units so that distance between neighboring sites is unity and $\vec{k}$ is in units of $|\vec{d_s} - \vec{d_r}|^{-1}$.

Applying the Bloch ansatz to \eq{eq:onsite_energy_general} we get,
\begin{equation}
\sum_s{H_{rs}\exp[i\vec{k}\cdot(\vec{d_s}-\vec{d_r})] \vert \phi,r \rangle} = E_r \vert \phi,r \rangle~,
\label{eq:datta_1}
\end{equation}
which for the 1D lattice becomes,
\begin{equation}
	[H_{r,r} + 2 H_{r,r+1}\cos(k)] \vert \phi,r \rangle = E_r \vert \phi,r \rangle~,
\label{eq:Bloch_1D}
\end{equation}
where $H_{r,r} = H^{\rm{JC}}_r - \mu N_r$ is the onsite Hamiltonian and $H_{r,r+1} = - \kappa(a_{r} a_{r+1}^\dagger + a_{r+1} a_{r}^\dagger)$ is the hopping Hamiltonian. As we have assumed identical sites, the chemical potential $\mu$ and hopping potential $\kappa$ are site independent and constant.

The $(2n \times 2n)$ matrix of \eq{eq:Bloch_1D} is block diagonal, and each $(2 \times 2)$ diagonal block is the same function of $n$. The eigenenergy of the extra polariton is ($n\geq0$),
\begin{equation}
\begin{split}
	E^p_{\pm}(n,k) &= \omega-\mu + \chi(n) - h(n,k)\\
		&\quad \pm \sqrt{h^2(n,k) - g(n,k) + \chi(n+1)^2},
\end{split}
\label{eq:eigenenergies_pol}
\end{equation}
\noindent where,
\begin{align}
	h(n,k) &\equiv [n + c_g^2(n)]\kappa\cos(k)~,\\
	g(n,k) &\equiv \{4(n+1)\sqrt{n}\beta c_e(n) c_g(n)\nonumber\\
		&\quad + [(n+1)c_g(n)^2-n c_e(n)^2]\Delta\} \kappa \cos(k)~.
\label{eq:h_g}
\end{align}
The energy of the hole is ($n>0$),
\begin{equation}
\begin{split}
	E^h_{\pm}(n,k) &= \mu - \omega + \chi(n) - h'(n,k)\\
		&\quad \pm \sqrt{h'^2(n,k) - g'(n,k) + \chi(n-1)^2},
\end{split}
\label{eq:eigenenergies_hol}
\end{equation}
\noindent where,
\begin{align}
	h'(n,k) &\equiv [n-c_e^2(n)]\kappa\cos(k)~,\\
	g'(n,k) &\equiv \{4(n-1)\sqrt{n}\beta c_e(n) c_g(n)\\
		&\quad + [(n c_g(n)^2-(n-1) c_e(n)^2]\Delta\} \kappa \cos(k)~.\nonumber
\label{eq:h_g_prime}
\end{align}

\begin{figure}[tb!]
		\includegraphics[width=80mm]{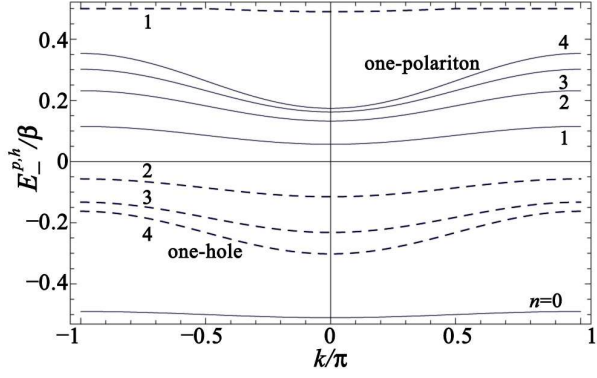}
\caption{The solid (dashed) lines are the one polariton (hole) band structures for $\Delta = 0$, $(\mu-\omega)/\beta=-0.5$, $\kappa/\beta=0.01$. The numbers indicate the $n$ value.}
\label{fig:ph_band}
\end{figure}

\fig{fig:ph_band} shows the one-polariton(hole) band structure. It forms the allowed energies a polariton can take near the Mott phase, analogous with conventional band structures which give the allowed energies an electron can take. 

\begin{figure}[tb!]
		\includegraphics[width=80mm]{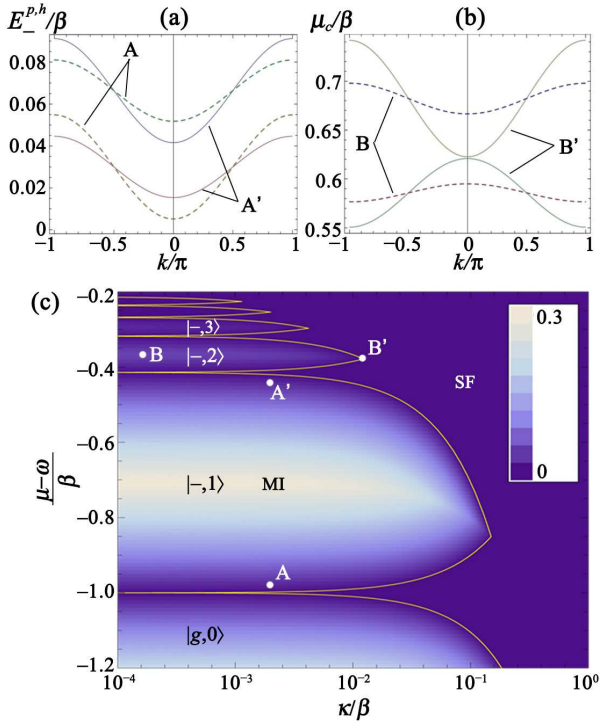}
\caption{(Color online)  (a) The dashed (solid) lines plot the one polariton-hole band structure at point $A$ ($A'$). At $A$ ($A'$) the energy cost of adding a hole is less (more) than the polariton counterpart. (b) The dashed (solid) lines plot $\mu_c^{p,h}$ at point $B$ ($B'$). The gap at $k=0$ gives the width of the Mott lobes. (c) Plot of the energy gap indicating the Mott lobes. Overlayed is the phase transition boundary. Indicated also are the Mott lobe states.}
\label{fig:phase_diagram}
\end{figure}

An energy gap at zero temperature indicates the presence of a MI phase. When the energy gap is zero, it costs no energy to add a polariton or hole, and we are no longer in the MI phase. This transition border occurs at critical chemical potential $\mu_c^p(n,0) = E^p_-(n,0) + \mu$ and $\mu_c^h(n,0) = E^h_-(n,0) - \mu$. $\mu_c^p(n,0)$ will form the upper boundary of the $n$th Mott lobe, and $\mu_c^h(n,0)$ will form its lower boundary. The exception is $n=0$, where there is no lower boundary. \fig{fig:phase_diagram}(a)-(c) shows the band structure and critical chemical potential spectrum at various points, and phase transition diagram. Overlayed in \fig{fig:phase_diagram}(c) is the aforementioned phase transition boundary.

\fig{fig:phase_diagram}(a) gives the polariton and hole energy bands for point $A$ and $A'$ in \fig{fig:phase_diagram}(c). Just above the curve corresponding to $\mu_c^h$ (e.g. point $A$) the energy cost of adding a hole is less than that of adding a polariton. Just below the curve corresponding to $\mu_c^p$ (e.g. point $A'$) the cost of adding a polariton is less than that of a hole. These minimal costs are the energy band gaps. \fig{fig:phase_diagram}(c) plots the energy gap forming the Mott lobes. 

The size of the gap indicates the level of stability, with larger gaps being more stable. The widths of the Mott lobes are given by $\mu_c^p(n,0) - \mu_c^h(n,0)$. \fig{fig:phase_diagram}(b) plots $\mu_c^{p,h}(n,k)$ spectrum near and far from the lobe tip. 

\begin{figure}[tb!]
	\includegraphics[width=80mm]{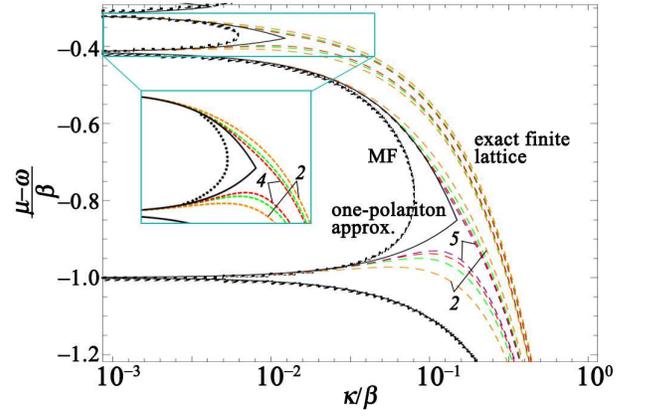}
	\caption{(Color online)  Comparison of one-polariton approximation (solid line), MF (dotted line), and finite cavity simulations (colored dashed lines). One mean excitation plateaux boundary are calculated for 2 to 5 cavities. Two mean excitiation plateaux boundary for 2 to 4 cavities. As the number of cavities increases it appears that finite cavity simulations tend to one-polariton analysis and not MF. }
\label{fig:phase_plot_comparison}
\end{figure}

As a means of comparison we overlay the one-polarition (hole) approximation phase transition border with that determined by a MF analysis in \fig{fig:phase_plot_comparison}. In the MF analysis the decoupling approximation $a_s^\dagger a_r = \langle a_r^\dagger \rangle a_r + \langle a_s \rangle a_s^\dagger - \langle a_s \rangle \langle a_r^\dagger \rangle$ is made, so that the JCH Hamiltonian is expressed as a sum over single sites,
\begin{equation}
	H^{\rm{MF}} = \sum_s{[H_s^{\rm{JC}} - z\kappa\psi(a_s^\dagger + a_s) + z\kappa|\psi|^2 - \mu N_s]},
\label{eq:jch_mf}
\end{equation}
where $\psi \equiv \langle a \rangle$ is known as the SF order parameter. It has been proposed as a measure of superfluidity \cite{vanOosten01}, where $\psi=0$ indicates MI, and $\psi>0$ indicates superfluidity. $z$ is the number of neighbors.

In Ref.~\cite{makin07} it was shown that the lobes in the MF approximation correlate well with the plateaux of mean excitations seen in finite cavity systems, demonstrating the progression from the finite case to the infinite. \fig{fig:phase_plot_comparison} shows the connection between these finite calculations compared with the MF and one-polariton approximation about the first and second Mott lobe. As the number of cavities increases we see that the exact results for the quantum phase transition points tend towards the one-polariton approximation and not the MF approximation. This highlights the physics missed in the MF analysis, but captured in our treatment, namely the spatial superposition states of the single polariton.

\section{semiconductor simulator}
\label{sec:Semi-Fluid In One Dimension}

The solid light system with tunable onsite and inter-site coupling parameters and extensions to 2D and 3D is able in principle to capture the physics of many non-trivial condensed matter systems. As an introductory example we consider a case of the solid light system inspired by conventional semiconductor doping.

Semiconductors have narrow energy band gaps. A material may be intrinsically or extrinsically semiconductive. Silicon, germanium, and gallium-arsenide are examples of intrinsic semiconductors. Extrinsic semiconductors are doped with impurities. These impurities introduce allowed energy states within the band gaps of the host material. The introduced energy band gap, known as dopant-site bonding energy, is characteristically small so that it takes little energy to ionize the dopant atoms and create free carriers in the conduction or valence bands. For example doping silicon with arsenic will create free electrons in the conduction band (n-type semiconductor). Doping with boron will create positive hole carriers in the valence band (p-type semiconductor). A typical level of impurity concentration would for example be 1 boron atom to $10^5$ silicon atoms.

\begin{figure}[tb!]
		\includegraphics[width=80mm]{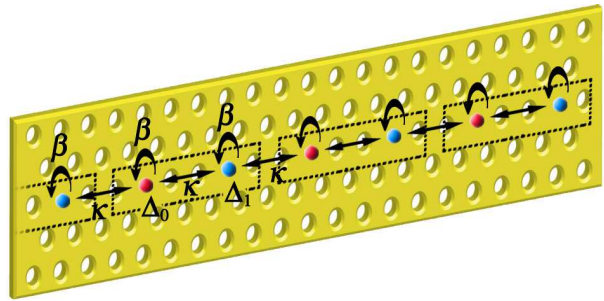}
\caption{(Color online) Every second atomic site is detuned from, but not far from, cavity resonance ($\Delta_0 = 0$, $\Delta_1=\beta$). The unit cell (dotted box) constitutes two adjacent sites. This introduces extra energy bands within the energy gap of the tuned system, in analogy to the semiconductor doping process.}
\label{fig:half_concentration}
\end{figure}

Here we show how this property of semiconductors can be captured in a solid light system. Motivated by the semiconductor doping process, our analog process simply involves changing the atomic transition of some of the two level atoms of the JC cavities. A tunable way of doing this is by Stark shift. We consider the 1D case where every second site is detuned, but not far from resonance as shown in \fig{fig:half_concentration}. Although the 1:1 concentration level is explicitly treated here, the physical properties which we show here are general to other doping concentrations. Embedding of a two-level system in waveguides \cite{Shen07} and in a single resonator in coupled-resonator waveguides have been proposed for controlled scattering of photons \cite{Zhou08}. Our model differs here in that all our cavities are embedded with two-level atoms, and we are controlling the detuning of multiple cavities.

The unit cell, i.e. the minimum site grouping to form a periodic lattice, now constitutes two adjacent sites. Thus we need to extend the Bloch state of \eq{eq:discrete_bloch} to a two site basis, 
\begin{equation}
	\vert \phi,s \rangle \otimes \vert \phi,s+1 \rangle = \vert \phi,r \rangle \otimes \vert \phi,r+1 \rangle \exp[i k(d_s-d_r)],
\label{eq:discrete_bloch_2}
\end{equation}
and the onsite Hamiltonian becomes, 
\begin{equation}
	H_{r,r} = \sum_{r=1}^{2}(H^{\rm{JC}}_r - \mu N_r) - \kappa(a_1^\dagger a_{2} + a_1 a_{2}^\dagger).
\label{eq:onsite_hamiltonian_2}
\end{equation}
The hopping Hamiltonian does not change.

\begin{figure*}[tb!]
		\includegraphics[width=160mm]{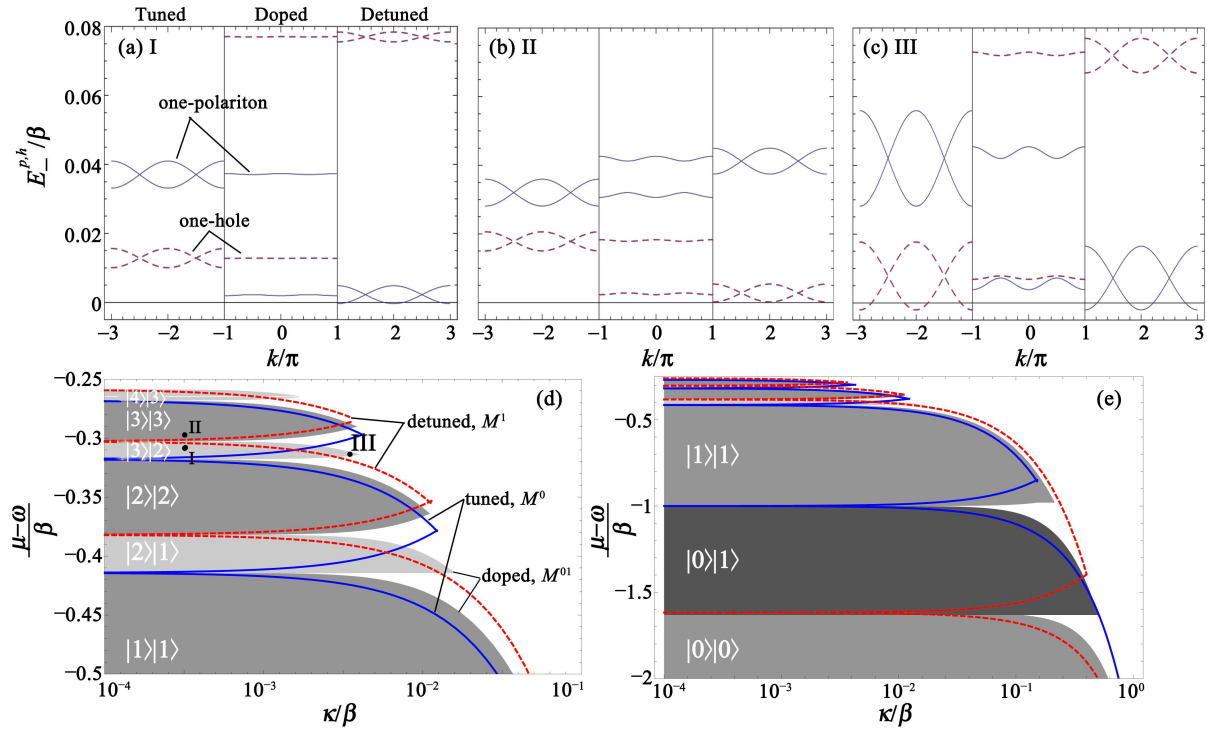}
\caption{(Color online)  (a) The solid (dashed) lines are the polariton (hole) band structure at point I. The left/middle/right panel is the band structure for the tuned/doped/detuned systen respectively. The impurity has introduced extra energy bands in the doped system reducing the energy band gap. The lowest energy band in the doped system is that of an extra polariton. This is analogous to a n-type semiconductor. (b) The solid (dashed) lines are the polariton (hole) band structure at point II. The lowest energy band in the doped system is that of a hole. This is analogous to a p-type semiconductor. (c) The solid (dashed) lines are the polariton (hole) band structure at point III. Interestingly the impurity increases the band gap: this has no semiconductor analogy. (d) The filled regions are the $\vert n \rangle_{\Delta_0} \otimes \vert n \rangle_{\Delta_1}$ (grey) and $\vert n + 1 \rangle_{\Delta_0} \otimes \vert n \rangle_{\Delta_1}$ (light grey) Mott lobes of the doped system. (e) The filled regions are the $\vert n \rangle_{\Delta_0} \otimes \vert n \rangle_{\Delta_1}$ (grey) and $\vert n - 1 \rangle_{\Delta_0} \otimes \vert n \rangle_{\Delta_1}$ (dark grey) Mott lobes of the doped system. Overlayed in (d) and (e) in solid (dashed) lines is the phase transition border for the tuned (detuned) system. Note the different scales of (d) and (e). For notational convenience $\vert n \rangle_{\Delta_0} \otimes \vert m \rangle_{\Delta_1}$ is written as $\vert n\rangle \vert m\rangle$.}
\label{fig:configuration_and_band_structure_50_a}
\end{figure*}

The Hamiltonian of \eq{eq:Bloch_1D} will now be a $(4n \times 4n)$ block diagonal matrix, where the diagonal blocks are $(4 \times 4)$ matrices. As before the blocks are the same function of $n$. Although an analytical solution is straightforward to obtain, it involves many terms and yields no insight, so we do not state it here. Instead we use it to plot the polariton-hole band structure and Mott lobes in \fig{fig:configuration_and_band_structure_50_a} for tuned ($\Delta_0 = \Delta_1 = 0$), doped ($\Delta_0 = 0, \Delta_1 = \beta$), and detuned ($\Delta_0 = \Delta_1 = \beta$) systems at different points of the phase diagram. 

\fig{fig:configuration_and_band_structure_50_a}(a) shows the polariton-hole band structure for point I in \fig{fig:configuration_and_band_structure_50_a}(d). The doped system has an extra polariton (hole) band. The extra band reduces the band gap from the tuned system. Here the lowest energy band is that of an extra polariton. This is analogous to the n-type semiconductor. Contrastingly \fig{fig:configuration_and_band_structure_50_a}(b) shows the polariton-hole band structure for point II in \fig{fig:configuration_and_band_structure_50_a}(d), where the lowest energy band is that of a hole. This is analogous to the p-type semiconductor.

Interestingly the solid light system also exhibits a feature with no semiconductor analog. \fig{fig:configuration_and_band_structure_50_a}(c) shows the polariton-hole band structure for point III in \fig{fig:configuration_and_band_structure_50_a}(a). Here the doped system \emph{increases} the band gap from \emph{both} the tuned and detuned system, extending the MI phase of the doped system into the SF region of the tuned and detuned systems. This new effect speaks to the richness of the JCH systems and should open new avenues for investigation.

The filled regions in \fig{fig:configuration_and_band_structure_50_a}(d)-(e) are the Mott lobes of the doped system. Because the detuning is not far from resonance, the Mott lobe states are: $\vert n \rangle_{\Delta_0} \otimes \vert n \rangle_{\Delta_1}$,$\vert n \pm 1 \rangle_{\Delta_0} \otimes \vert n \rangle_{\Delta_1}$, where $\vert n \rangle_{\Delta_0}$ is the JC eigenstate $\vert-,n\rangle$ with $\Delta_0$ detuning.

For a given $n$, the Mott lobe states will be,
\begin{eqnarray}
	|n\rangle_{\Delta_0} \otimes |n\rangle_{\Delta_1}, |n+1\rangle_{\Delta_0} \otimes |n\rangle_{\Delta_1},~~~~~~~~~~~~~~~~~~~~\nonumber\\
	{\rm if}~\mu_c(n,\Delta_0) < \mu_c(n,\Delta_1),\\
	|n\rangle_{\Delta_0} \otimes |n\rangle_{\Delta_1}, |n\rangle_{\Delta_0} \otimes |n+1\rangle_{\Delta_1},~~~~~~~~~~~~~~~~~~~~\nonumber\\
	{\rm if}~\mu_c(n,\Delta_0) > \mu_c(n,\Delta_1),	
	\label{eq:MI_form_general}
\end{eqnarray}

In \fig{fig:configuration_and_band_structure_50_a}(d)-(e) where $\Delta_0 = 0$ and $\Delta_1 = \beta$, the Mott lobe states are, 
\begin{eqnarray}
	|n\rangle_{\Delta_0} \otimes |n\rangle_{\Delta_1}, |n+1\rangle_{\Delta_0} \otimes |n\rangle_{\Delta_1}, & {\rm for}~n>0,\\
	|n\rangle_{\Delta_0} \otimes |n\rangle_{\Delta_1}, |n\rangle_{\Delta_0} \otimes |n+1\rangle_{\Delta_1}, & {\rm for}~n=0.
	\label{eq:eq:MI_form_example}
\end{eqnarray}

In \fig{fig:configuration_and_band_structure_50_a}(d)-(e), overlayed in solid (dashed) lines is the Mott borders of the tuned (detuned) system. For small hopping the set of Mott lobes of the doped system ($M^{01}$) is the intersection of the set of Mott lobes of the tuned ($M^{0}$) and detuned systems ($M^{1}$), $M^{01} = M^0 \cap M^1$, thus there are twice as many Mott lobes of smaller size in the doped system then in the tuned or detuned system. As hopping increases, the inter-cavity correlation causes divergence from this relation. Since $E^{p,h}_-(\mu_c) = 0$, $|\mu_c - \mu|_{\rm{min}}$ gives the band gap. Hence the reduction in band gap which makes the Mott phase less stable to thermal fluctuations in the doped system is directly indicated by the smaller Mott lobe widths.

\section{Conclusion}

We have used the one-polariton approximation to find the band structure of a solid light system. We have shown the one-polariton approximation to capture inter-cavity correlation effects missed by MF analysis. Using the band structures we have shown the phase transition diagram, highlighting levels of MI stability. We have further demonstrated the similarities between the fields of solid state and solid light by showing a solid light analogy of the extrinsic semiconductor. Controlled site detuning can be seen as analogous to semiconductor doping, and we find that this process reduces the energy band gap forming smaller Mott lobes, but unlike the semiconductor, it can also increase the energy band gap, extending the Mott lobe tip. These results open the way for the treatment of other condensed matter systems in this framework.

\section{Acknowledgments}

We acknowledge useful discussions with J.H. Cole and A. Hayward. A.D.G. and L.C.L.H. acknowledge the Australian Research Council for financial support Projects No. DP0880466 and No. DP0770715, respectively. This work was supported in part by the Australian Research Council, the Australian Government and by the U.S. National Security Agency, and the U.S. Army Research Office under Contract No. W911NF-08-1-0527.

\bibliography{band_structure_of_solid_light}

\end{document}